\RequirePackage{lineno}
\documentclass[aps,prd,twocolumn,superscriptaddress]{revtex4}
\usepackage{graphicx}

\begin{document}


\title{Measurement of the Top Quark Mass in the Lepton+Jets Channel Using
       the Lepton Transverse Momentum}


\affiliation{Institute of Physics, Academia Sinica, Taipei, Taiwan 11529, Republic of China} 
\affiliation{Argonne National Laboratory, Argonne, Illinois 60439, USA} 
\affiliation{University of Athens, 157 71 Athens, Greece} 
\affiliation{Institut de Fisica d'Altes Energies, ICREA, Universitat Autonoma de Barcelona, E-08193, Bellaterra (Barcelona), Spain} 
\affiliation{Baylor University, Waco, Texas 76798, USA} 
\affiliation{Istituto Nazionale di Fisica Nucleare Bologna, $^z$University of Bologna, I-40127 Bologna, Italy} 
\affiliation{University of California, Davis, Davis, California 95616, USA} 
\affiliation{University of California, Los Angeles, Los Angeles, California 90024, USA} 
\affiliation{Instituto de Fisica de Cantabria, CSIC-University of Cantabria, 39005 Santander, Spain} 
\affiliation{Carnegie Mellon University, Pittsburgh, Pennsylvania 15213, USA} 
\affiliation{Enrico Fermi Institute, University of Chicago, Chicago, Illinois 60637, USA}
\affiliation{Comenius University, 842 48 Bratislava, Slovakia; Institute of Experimental Physics, 040 01 Kosice, Slovakia} 
\affiliation{Joint Institute for Nuclear Research, RU-141980 Dubna, Russia} 
\affiliation{Duke University, Durham, North Carolina 27708, USA} 
\affiliation{Fermi National Accelerator Laboratory, Batavia, Illinois 60510, USA} 
\affiliation{University of Florida, Gainesville, Florida 32611, USA} 
\affiliation{Laboratori Nazionali di Frascati, Istituto Nazionale di Fisica Nucleare, I-00044 Frascati, Italy} 
\affiliation{University of Geneva, CH-1211 Geneva 4, Switzerland} 
\affiliation{Glasgow University, Glasgow G12 8QQ, United Kingdom} 
\affiliation{Harvard University, Cambridge, Massachusetts 02138, USA} 
\affiliation{Division of High Energy Physics, Department of Physics, University of Helsinki and Helsinki Institute of Physics, FIN-00014, Helsinki, Finland} 
\affiliation{University of Illinois, Urbana, Illinois 61801, USA} 
\affiliation{The Johns Hopkins University, Baltimore, Maryland 21218, USA} 
\affiliation{Institut f\"{u}r Experimentelle Kernphysik, Karlsruhe Institute of Technology, D-76131 Karlsruhe, Germany} 
\affiliation{Center for High Energy Physics: Kyungpook National University, Daegu 702-701, Korea; Seoul National University, Seoul 151-742, Korea; Sungkyunkwan University, Suwon 440-746, Korea; Korea Institute of Science and Technology Information, Daejeon 305-806, Korea; Chonnam National University, Gwangju 500-757, Korea; Chonbuk National University, Jeonju 561-756, Korea} 
\affiliation{Ernest Orlando Lawrence Berkeley National Laboratory, Berkeley, California 94720, USA} 
\affiliation{University of Liverpool, Liverpool L69 7ZE, United Kingdom} 
\affiliation{University College London, London WC1E 6BT, United Kingdom} 
\affiliation{Centro de Investigaciones Energeticas Medioambientales y Tecnologicas, E-28040 Madrid, Spain} 
\affiliation{Massachusetts Institute of Technology, Cambridge, Massachusetts 02139, USA} 
\affiliation{Institute of Particle Physics: McGill University, Montr\'{e}al, Qu\'{e}bec, Canada H3A~2T8; Simon Fraser University, Burnaby, British Columbia, Canada V5A~1S6; University of Toronto, Toronto, Ontario, Canada M5S~1A7; and TRIUMF, Vancouver, British Columbia, Canada V6T~2A3} 
\affiliation{University of Michigan, Ann Arbor, Michigan 48109, USA} 
\affiliation{Michigan State University, East Lansing, Michigan 48824, USA}
\affiliation{Institution for Theoretical and Experimental Physics, ITEP, Moscow 117259, Russia}
\affiliation{University of New Mexico, Albuquerque, New Mexico 87131, USA} 
\affiliation{Northwestern University, Evanston, Illinois 60208, USA} 
\affiliation{The Ohio State University, Columbus, Ohio 43210, USA} 
\affiliation{Okayama University, Okayama 700-8530, Japan} 
\affiliation{Osaka City University, Osaka 588, Japan} 
\affiliation{University of Oxford, Oxford OX1 3RH, United Kingdom} 
\affiliation{Istituto Nazionale di Fisica Nucleare, Sezione di Padova-Trento, $^{aa}$University of Padova, I-35131 Padova, Italy} 
\affiliation{LPNHE, Universite Pierre et Marie Curie/IN2P3-CNRS, UMR7585, Paris, F-75252 France} 
\affiliation{University of Pennsylvania, Philadelphia, Pennsylvania 19104, USA}
\affiliation{Istituto Nazionale di Fisica Nucleare Pisa, $^{bb}$University of Pisa, $^{cc}$University of Siena and $^{dd}$Scuola Normale Superiore, I-56127 Pisa, Italy} 
\affiliation{University of Pittsburgh, Pittsburgh, Pennsylvania 15260, USA} 
\affiliation{Purdue University, West Lafayette, Indiana 47907, USA} 
\affiliation{University of Rochester, Rochester, New York 14627, USA} 
\affiliation{The Rockefeller University, New York, New York 10065, USA} 
\affiliation{Istituto Nazionale di Fisica Nucleare, Sezione di Roma 1, $^{ee}$Sapienza Universit\`{a} di Roma, I-00185 Roma, Italy} 

\affiliation{Rutgers University, Piscataway, New Jersey 08855, USA} 
\affiliation{Texas A\&M University, College Station, Texas 77843, USA} 
\affiliation{Istituto Nazionale di Fisica Nucleare Trieste/Udine, I-34100 Trieste, $^{ff}$University of Trieste/Udine, I-33100 Udine, Italy} 
\affiliation{University of Tsukuba, Tsukuba, Ibaraki 305, Japan} 
\affiliation{Tufts University, Medford, Massachusetts 02155, USA} 
\affiliation{University of Virginia, Charlottesville, VA  22906, USA}
\affiliation{Waseda University, Tokyo 169, Japan} 
\affiliation{Wayne State University, Detroit, Michigan 48201, USA} 
\affiliation{University of Wisconsin, Madison, Wisconsin 53706, USA} 
\affiliation{Yale University, New Haven, Connecticut 06520, USA} 
\author{T.~Aaltonen}
\affiliation{Division of High Energy Physics, Department of Physics, University of Helsinki and Helsinki Institute of Physics, FIN-00014, Helsinki, Finland}
\author{B.~\'{A}lvarez~Gonz\'{a}lez$^v$}
\affiliation{Instituto de Fisica de Cantabria, CSIC-University of Cantabria, 39005 Santander, Spain}
\author{S.~Amerio}
\affiliation{Istituto Nazionale di Fisica Nucleare, Sezione di Padova-Trento, $^{aa}$University of Padova, I-35131 Padova, Italy} 

\author{D.~Amidei}
\affiliation{University of Michigan, Ann Arbor, Michigan 48109, USA}
\author{A.~Anastassov}
\affiliation{Northwestern University, Evanston, Illinois 60208, USA}
\author{A.~Annovi}
\affiliation{Laboratori Nazionali di Frascati, Istituto Nazionale di Fisica Nucleare, I-00044 Frascati, Italy}
\author{J.~Antos}
\affiliation{Comenius University, 842 48 Bratislava, Slovakia; Institute of Experimental Physics, 040 01 Kosice, Slovakia}
\author{G.~Apollinari}
\affiliation{Fermi National Accelerator Laboratory, Batavia, Illinois 60510, USA}
\author{J.A.~Appel}
\affiliation{Fermi National Accelerator Laboratory, Batavia, Illinois 60510, USA}
\author{A.~Apresyan}
\affiliation{Purdue University, West Lafayette, Indiana 47907, USA}
% ...
\collaboration{CDF Collaboration\footnote{With visitors from $^a$University of Massachusetts Amherst, Amherst, Massachusetts 01003,
$^b$Istituto Nazionale di Fisica Nucleare, Sezione di Cagliari, 09042 Monserrato (Cagliari), Italy,
$^c$University of California Irvine, Irvine, CA  92697, 
$^d$University of California Santa Barbara, Santa Barbara, CA 93106
$^e$University of California Santa Cruz, Santa Cruz, CA  95064,
$^f$CERN,CH-1211 Geneva, Switzerland,
$^g$Cornell University, Ithaca, NY  14853, 
$^h$University of Cyprus, Nicosia CY-1678, Cyprus, 
$^i$University College Dublin, Dublin 4, Ireland,
$^j$University of Fukui, Fukui City, Fukui Prefecture, Japan 910-0017,
$^k$Universidad Iberoamericana, Mexico D.F., Mexico,
$^l$Iowa State University, Ames, IA  50011,
$^m$University of Iowa, Iowa City, IA  52242,
$^n$Kinki University, Higashi-Osaka City, Japan 577-8502,
$^o$Kansas State University, Manhattan, KS 66506,
$^p$University of Manchester, Manchester M13 9PL, England,
$^q$Queen Mary, University of London, London, E1 4NS, England,
$^r$Muons, Inc., Batavia, IL 60510,
$^s$Nagasaki Institute of Applied Science, Nagasaki, Japan, 
$^t$National Research Nuclear University, Moscow, Russia,
$^u$University of Notre Dame, Notre Dame, IN 46556,
$^v$Universidad de Oviedo, E-33007 Oviedo, Spain, 
$^w$Texas Tech University, Lubbock, TX  79609, 
$^x$Universidad Tecnica Federico Santa Maria, 110v Valparaiso, Chile,
$^y$Yarmouk University, Irbid 211-63, Jordan,
$^{gg}$On leave from J.~Stefan Institute, Ljubljana, Slovenia, 
}}
\noaffiliation


\date{\today}

\begin{abstract}
This letter reports a measurement of the top quark mass, $M_{top}$, in
data from $p\bar{p}$ collisions at $\sqrt{s}=1.96$ TeV corresponding to
2.7 fb$^{-1}$ of integrated luminosity at the Fermilab Tevatron using the
CDF II detector. Events with the lepton+jets topology are selected. An
unbinned likelihood is constructed based on the dependence of the lepton
transverse momentum, $P_{\rm T}$, on $M_{top}$. A maximum likelihood fit
to the data yields a measured mass
$M_{top}=176.9\pm 8.0_{\rm stat}\pm 2.7_{\rm syst}$ GeV/$c^{2}$. In this
measurement, the contribution by the jet energy scale uncertainty to the
systematic error is negligible. The result provides an important consistency
test for other $M_{top}$ measurements where explicit use of the jet energy
is made for deriving the top quark mass.
\end{abstract}


\maketitle



\section{Introduction}
\label{sec:introduction}

The top quark is the heaviest known fundamental particle. Since the
discovery of this particle in 1995 at the Fermilab Tevatron \cite{top},
both the CDF and D0 experiments have been improving
the precision of the measurement of its mass $M_{top}$, which is a fundamental
parameter in the Standard Model (SM) of particle physics. Loop corrections
in electroweak theory relate $M_{top}$ and the $W$ boson mass $M_{W}$ to the
mass of the predicted Higgs boson \cite{mtop}. Therefore, precision
measurements of $M_{top}$ provide constraints on the value of the Higgs
boson mass as well as a consistency check of the SM electroweak theory
\cite{higgs}.

The largest systematic uncertainties in the measurement of $M_{top}$
are due to uncertainties in the measurement of jets. Jets are composite
objects which must be associated with the partons produced in $t\bar{t}$
decays using jet-parton combinatorics and energy transfer functions derived
from Monte Carlo (MC) simulation. Measuring the jet energy requires detailed
corrections and an overall scale calibration. The {\it in situ} energy scale
calibration with a $W$ mass constraint used in other $M_{top}$ measurement
techniques \cite{mtm2} is not directly applicable to jets produced by $b$
quarks. On the other hand, charged leptons (electrons or muons) produced in
$t\bar{t}$ decays are directly observable in the detector and their momenta
can be measured with very high precision. Leptons thus provide a very clean
probe of the kinematics of $t\bar{t}$ decays. The sensitivity of their
momentum on the top quark mass can be used to measure $M_{top}$ without the
complexities and related uncertainties which are inherent to the use of jets,
albeit with less statistical precision. The result reported in this letter is,
therefore, complementary to the existing precision measurements by having
different systematic uncertainties than the previously published results.

In $p\bar{p}$ collisions, top quarks are produced predominantly as
$t\bar{t}$ pairs. Within the SM, the top quark decays almost exclusively
into a $W$ boson and a bottom quark $b$ \cite{pdg}. The events in which
one of the $W$ bosons decays leptonically to a charged lepton and a neutrino
and the other decays hadronically into two jets define the ``lepton+jets''
decay channel,
$t\bar{t}\to W^{+}bW^{-}\bar{b}\to l\nu_{l}bq\bar{q}^{\prime}\bar{b}$.
Including the two jets arising from the $b$ quarks, the lepton+jets
topology contains at least four jets. Additional jets may be generated
from gluon radiation or from soft hadron interactions forming the
``underlying event''. Despite this complication, the lepton+jets channel
provides the best balance of available statistics and sample purity.

The sensitivity of kinematic variables of the lepton to $M_{top}$ has to
be studied using a MC model of $t\bar{t}$ events. In a preliminary study
of the possibility of using only leptons to measure $M_{top}$ it was found
that the optimum variable to use is the lepton transverse momentum $P_{\rm T}$
\cite{lepPt}. This variable is generically used to signify the transverse
momentum of muons measured in the tracker or the transverse energy of
electrons measured in the calorimeter \cite{frame}, which offers better
resolution than the tracker for high energy electrons.

The only previous analysis using the transverse momentum of leptons with
data of 1.9 fb$^{-1}$ integrated luminosity from the CDF experiment \cite{Lxy}
explored the mean value of the lepton $P_{\rm T}$. That analysis also exploited
the mean value of the transverse path length of b-flavored hadrons produced in
$t\bar{t}$ decays and combined the two results to measure $M_{top}$ in the
lepton+jets channel of $t\bar{t}$ decays. In the measurement reported in this
letter a shape analysis of the lepton $P_{\rm T}$ spectrum is applied in the
same decay channel. This technique is less sensitive to acceptance related
effects, which alter the mean value of the lepton $P_{\rm T}$ but leave the
shape of the $P_{\rm T}$ distribution unchanged, and thus allows for a
measurement with smaller systematic uncertainties.

\section{Detector and event selection}
\label{sec:selection}

The measurement described in this letter uses data collected with the
Collider Detector at Fermilab (CDF) II detector \cite{cdf} at the Tevatron
$p\bar{p}$ collider corresponding to an integrated luminosity of 2.7
fb$^{-1}$. CDF is a cylindrically symmetric detector surrounding the
colliding beams. It consists, radially from inside to outside, of an inner
silicon tracker allowing for accurate vertex reconstruction and an outer
wire chamber tracker, both operating in a uniform magnetic field of 1.4
Tesla which is produced by a superconducting solenoid surrounding the
tracker; scintillators for time-of-flight measurements; a sampling calorimeter
with an inner electromagnetic compartment and an outer hadronic compartment;
and wire chambers for muon identification. The tracking system measures
charged particle tracks with a transverse momentum precision of
$\Delta P_{\rm T}/P_{\rm T}^{2}=0.07\%({\rm GeV/}c)^{-1}$. The central
calorimeters have an electromagnetic (hadronic) energy resolution of
$\sigma(E_{\rm T})/E_{\rm T}\sim 13.5\%/\sqrt{E_{\rm T}({\rm GeV})}\oplus 1.5\%$
($\sigma(E_{\rm T})/E_{\rm T}\sim 50\%/\sqrt{E_{\rm T}({\rm GeV})}\oplus 3\%$)
and a tower segmentation of
$\Delta\eta\times\Delta\phi\simeq 0.1\times 15^{\circ}$.

\begin{table}
\caption{Expected and observed lepton + $\ge$4 jets sample for an integrated
luminosity of 2.7 fb$^{-1}$. The $t\bar{t}$ contribution is estimated using
a cross section of 6.7 pb \cite{xsec} and $M_{top}=175$ GeV/$c^{2}$. The
uncertainties are statistical only.
\label{tab:expSM}}
\begin{center}
\begin{tabular}{lc}
\hline\hline
Event type & Expected number of events \\ \hline
$W$+heavy flavor & 91.3$\pm$15.6 \\
$W$+light flavor & 29.6$\pm$6.0 \\
$Z\to ll+$jets & 5.3$\pm$1.3 \\
Dibosons ($WW$, $WZ$, $ZZ$) & 8.6$\pm$1.5 \\
Single top & 8.6$\pm$1.2 \\
Fakes & 34.9$\pm$22.5 \\
Total background & 178.8$\pm$28.2 \\
$t\bar{t}$ signal & 634.9$\pm$44.8 \\ \hline
Total expectation & 813.8$\pm$53.0 \\ \hline
Observed events & 858 \\ \hline\hline
\end{tabular}
\end{center}
\end{table}

The leptons used in this measurement were detected in the central region of
the CDF detector, covering a pseudorapidity range of
$\vert\eta\vert\le 1$ \cite{frame}, with an inclusive lepton trigger requiring
an electron with $E_{\rm T}\ge 18$ GeV or a muon with $P_{\rm T}\ge 18$
GeV/$c$. From this inclusive lepton dataset, events are selected offline in
the lepton+jets channel by requiring one electron with transverse energy
$E_{\rm T}\ge 20$ GeV or one muon with transverse momentum $P_{\rm T}\ge 20$
GeV/$c$, at least four jets with transverse energy $E_{\rm T}\ge 20$ GeV
and pseudorapidity $\vert\eta\vert\le 2$, and missing transverse energy
$\not\!\!{E}_{\rm T}\ge 20$ GeV \cite{met} to account for the unobserved
neutrino. Electrons are reconstructed as isolated energy clusters in the
electromagnetic calorimeter \cite{iso} and matched to tracks fiducial to
these clusters. Muons are reconstructed from tracks fiducial to the muon
chambers, matched to isolated tracks in the central tracker, and are required
to deposit minimal energy in the calorimeter. Jets are reconstructed as
energy clusters in the hadronic calorimeter towers within a cone of radius
0.4 \cite{iso} around a seed tower. Jet energies are corrected for
non-uniformities in the calorimeter response as a function of the jet
pseudorapidity, for multiple $p\bar{p}$ interactions in the event, and for
the energy scale of the calorimeter \cite{jets}. The expected fraction of
$t\bar{t}$ lepton+jets events passing the above selection requirements is
approximately 10\%. The signal to background fraction is S/B$\sim$0.5. To
enhance this fraction, at least one jet is required to be tagged as
originating from a heavy flavor quark using a secondary vertex tagging
algorithm \cite{btag}. The fraction of signal events passing this additional
requirement is reduced to $\sim$6\%, while S/B rises to $\sim$3.7. Decays of
the $W$ boson to a $\tau$ lepton which subsequently decays to an electron or
muon can also pass all selection requirements and they amount to approximately
7\% of the $t\bar{t}$ signal.

\begin{figure}
\includegraphics[width=8.66cm]{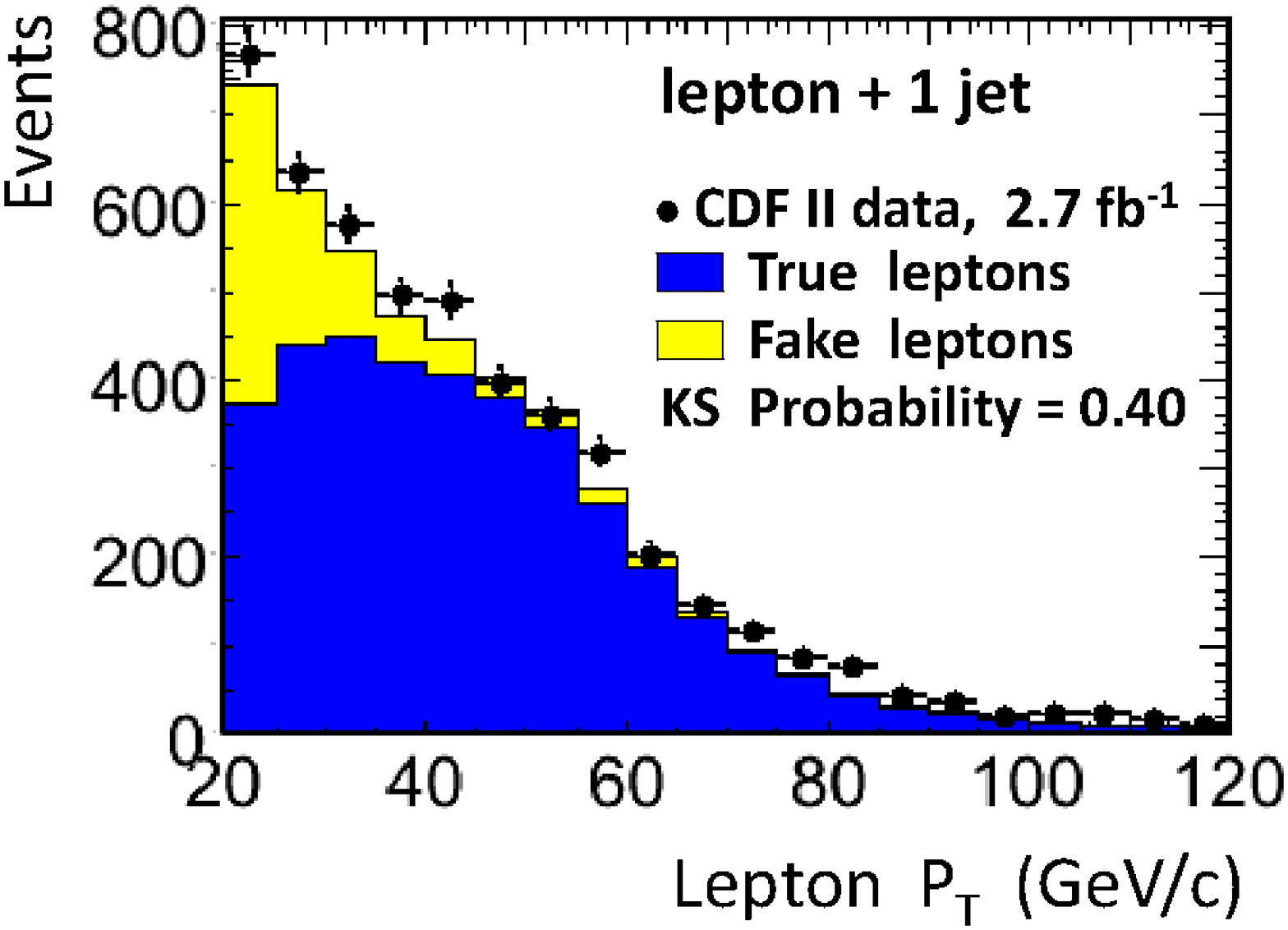}
\vspace{-0.2cm}
\includegraphics[width=8.66cm]{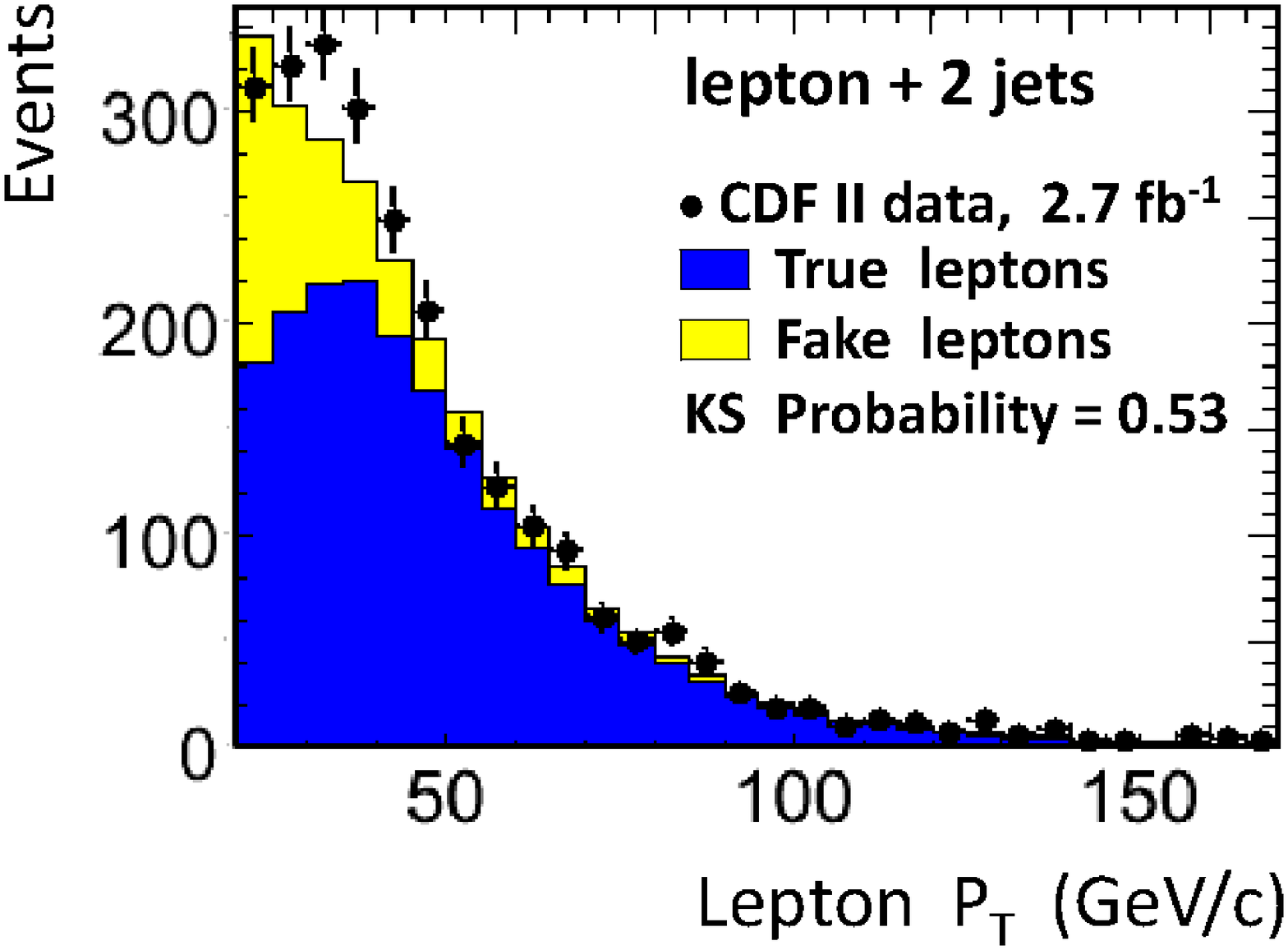}
\caption{Comparison of the expectations with the data in the lepton + 1 jet
and lepton + 2 jets samples.
\label{fig:bkg}}
\end{figure}

\section{Background estimation}
\label{sec:background}

Background events from other SM processes passing the selection criteria
contain: $W$ boson production associated with jets from heavy flavor quarks
($b\bar{b}$, $c\bar{c}$ or $c$); $W$ boson production associated with jets
from light flavor quarks; $Z$ boson production associated with jets
where the $Z$ decays leptonically and one lepton escapes detection, thus
giving rise to high $\not\!\!\!{E}_{\rm T}$; diboson events ($WW$, $WZ$,
$ZZ$) in which one boson decays leptonically and the other hadronically;
single-top events where the $W$ boson produced by the top quark decays
leptonically; and events having a jet misidentified as a lepton, subsequently
referred to as a ``fake'' lepton. The $t\bar{t}$ and diboson events were
modeled using the PYTHIA generator \cite{pythia}. The $W$+jets and $Z$+jets
events were modeled using the ALPGEN generator \cite{alpgen} with the parton
hadronization done by PYTHIA. The single top events were modeled using the
MADGRAPH/MADEVENT generator \cite{mad} with the parton hadronization done
again by PYTHIA. The modeled events were processed through the detector
and trigger simulation and reconstructed using the CDF II software
\cite{cdfsim}. The $P_{\rm T}$ distribution of fake leptons was modeled using
a data sample selected by requiring each event to fail at least one of the
criteria of good lepton selection. The complete sample composition was
estimated with a method used for the $t\bar{t}$ cross section measurement
\cite{xsec}. In this method the expected rates of $t\bar{t}$, $Z$+jets,
diboson and single-top events are estimated from the MC; the rate of
$W$+heavy-flavored jet events is estimated from the MC, adjusted to the
data in the 1- and 2-jet control bins using a neural network; the rate
of $W$+light-flavored jet events is estimated from the MC using a per-jet
probability of mis-tagging; and the rate of events with a fake lepton is
estimated by fitting the total MC + fake leptons $\not\!\!{E}_{\rm T}$
spectrum to the $\not\!\!{E}_{\rm T}$ spectrum of the data in the 4-jet bin
with the normalization of all other components fixed. More details on the
method can be found in \cite{xsec}.

\begin{figure}
\includegraphics[width=8.66cm]{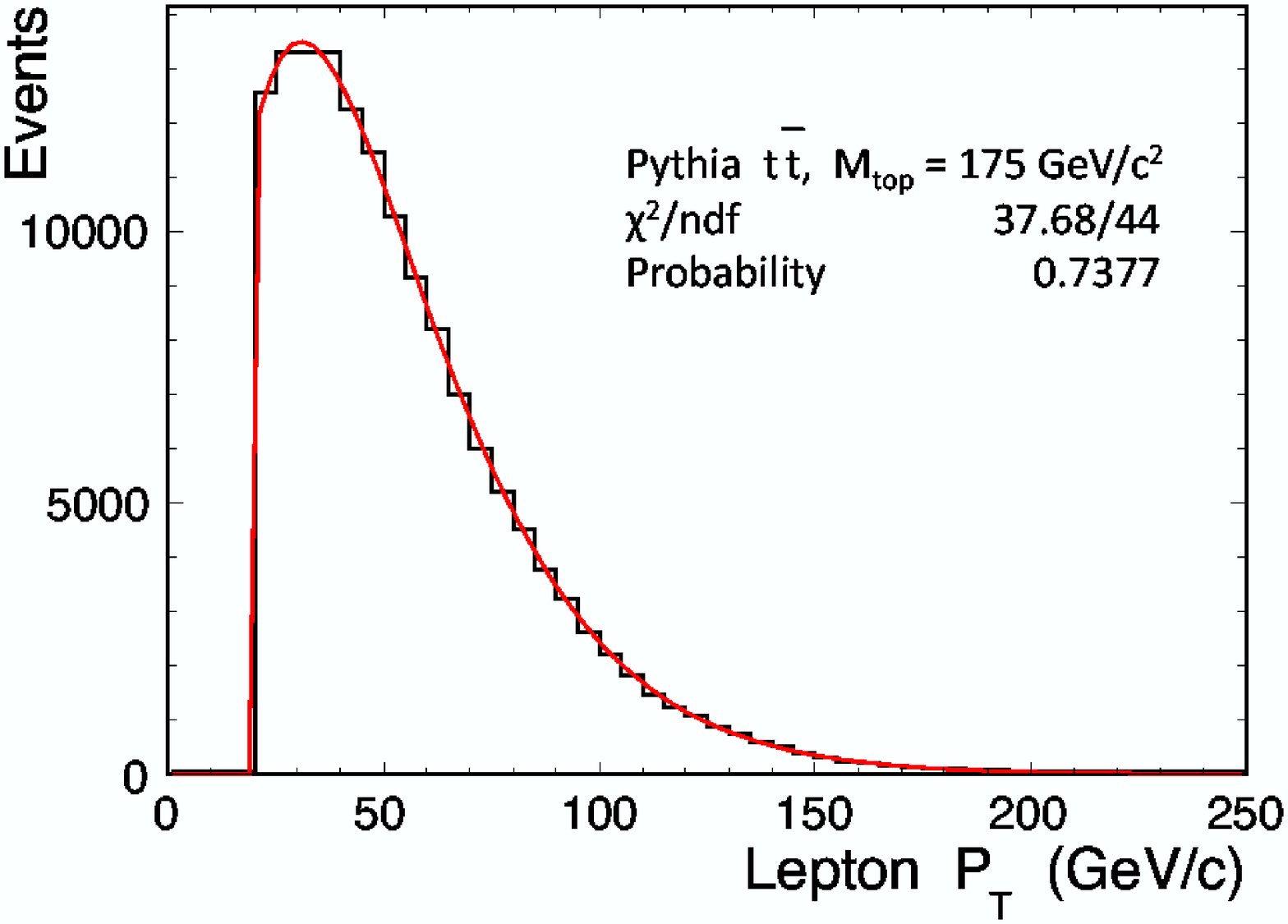}
\vspace{-0.2cm}
\includegraphics[width=8.66cm]{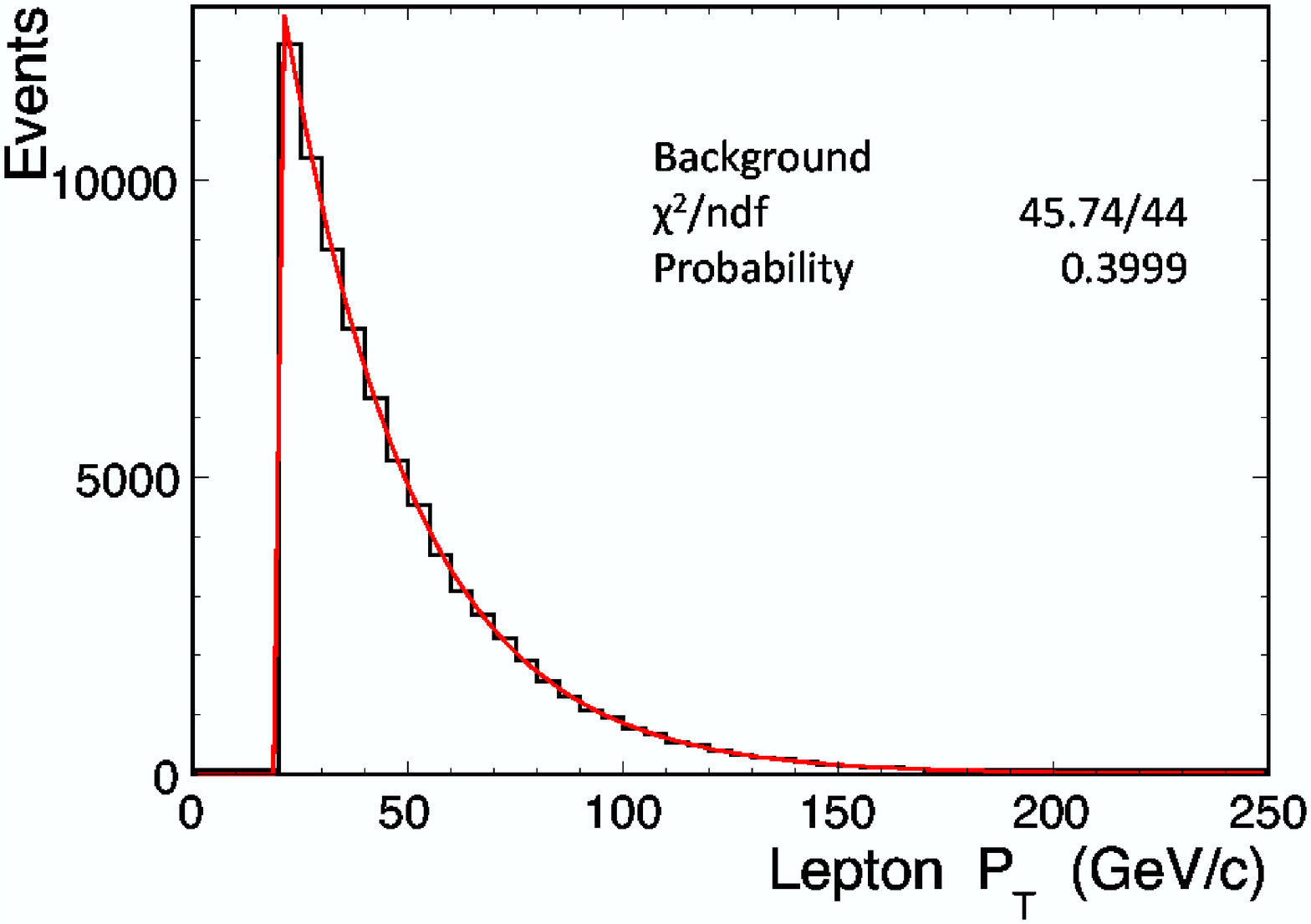}
\caption{{\bf Top:} The fit of the incomplete $\Gamma$ function to the
PYTHIA lepton $P_{\rm T}$ histogram with input $M_{top}=175$ GeV/$c^{2}$.
{\bf Bottom:} The fit of the incomplete $\Gamma$ function to the lepton
$P_{\rm T}$ histogram of the estimated total background.
\label{fig:fits}}
\end{figure}

The total expected composition of the selected events is shown in Table
\ref{tab:expSM}. Extensive validation tests of the background model were
made in the control samples of events with one lepton and one or two jets,
where the $t\bar{t}$ signal is expected to be negligible. Comparisons of
the expectations with the data in the control samples are shown in Figure
\ref{fig:bkg}. Only overall shape discrepancies are of concern for the
purpose of this measurement because the normalization of the total
background is varied in the fit to the data, as described in Section
\ref{sec:method}. There is a possible shape discrepancy between the total
expectation and the data in the lepton + 2 jets control sample for
$P_{\rm T}<50$ GeV/$c$, attributed to the fake lepton shape being inaccurate.
A systematic uncertainty from the fakes distribution is assigned to the
final $M_{top}$ measurement, as discussed in Section \ref{sec:result}.

\section{Method}
\label{sec:method}

\begin{table}
\caption{Fit parameters of the lepton $P_{\rm T}$ distribution model for the
$t\bar{t}$ signal $(\alpha_{1,2,3,4})$ and for the total background
$(\beta_{1,2})$.
\label{tab:coeff}}
\begin{center}
\begin{tabular}{llc}
\hline\hline
\multicolumn{2}{l}{Parameter} & Value \\ \hline
Intercept~~ & $\alpha_{1}$ & 1.72$\pm$0.10 \\
Slope & $\alpha_{2}$ [(GeV/$c$)$^{-1}$]~~~~~ & $-$0.0009$\pm$0.0004 \\
Intercept & $\alpha_{3}$ [GeV/$c$] & 6.19$\pm$0.26 \\
Slope & $\alpha_{4}$ & 0.079$\pm$0.001 \\ \hline
 & $\beta_{1}$ & 0.27$\pm$0.54 \\
 & $\beta_{2}$ [GeV/$c$] & 26.73$\pm$6.79 \\ \hline\hline
\end{tabular}
\end{center}
\end{table}

\begin{figure}
\includegraphics[width=8.66cm]{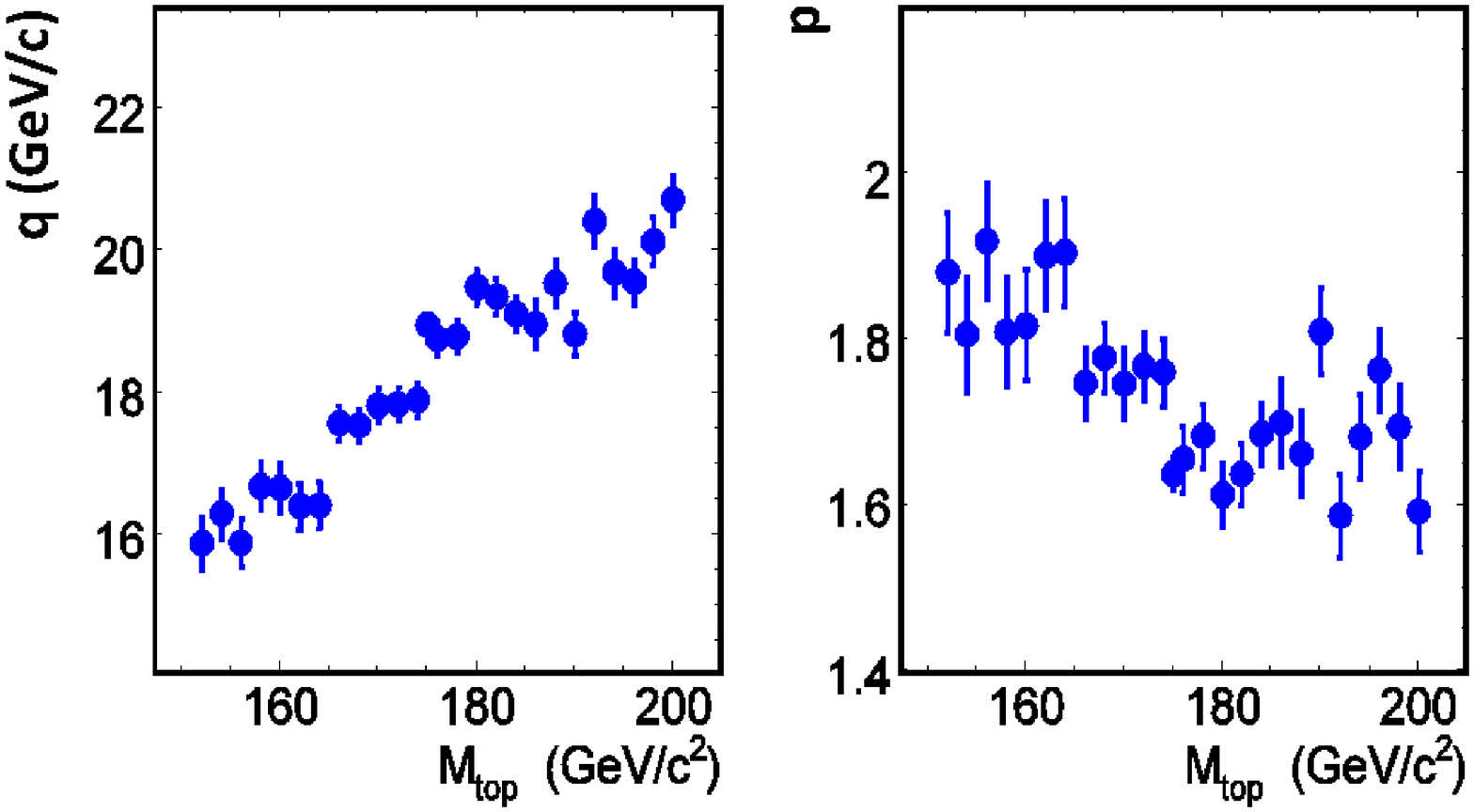}
\vspace{-1.2cm}
\includegraphics[width=8.66cm]{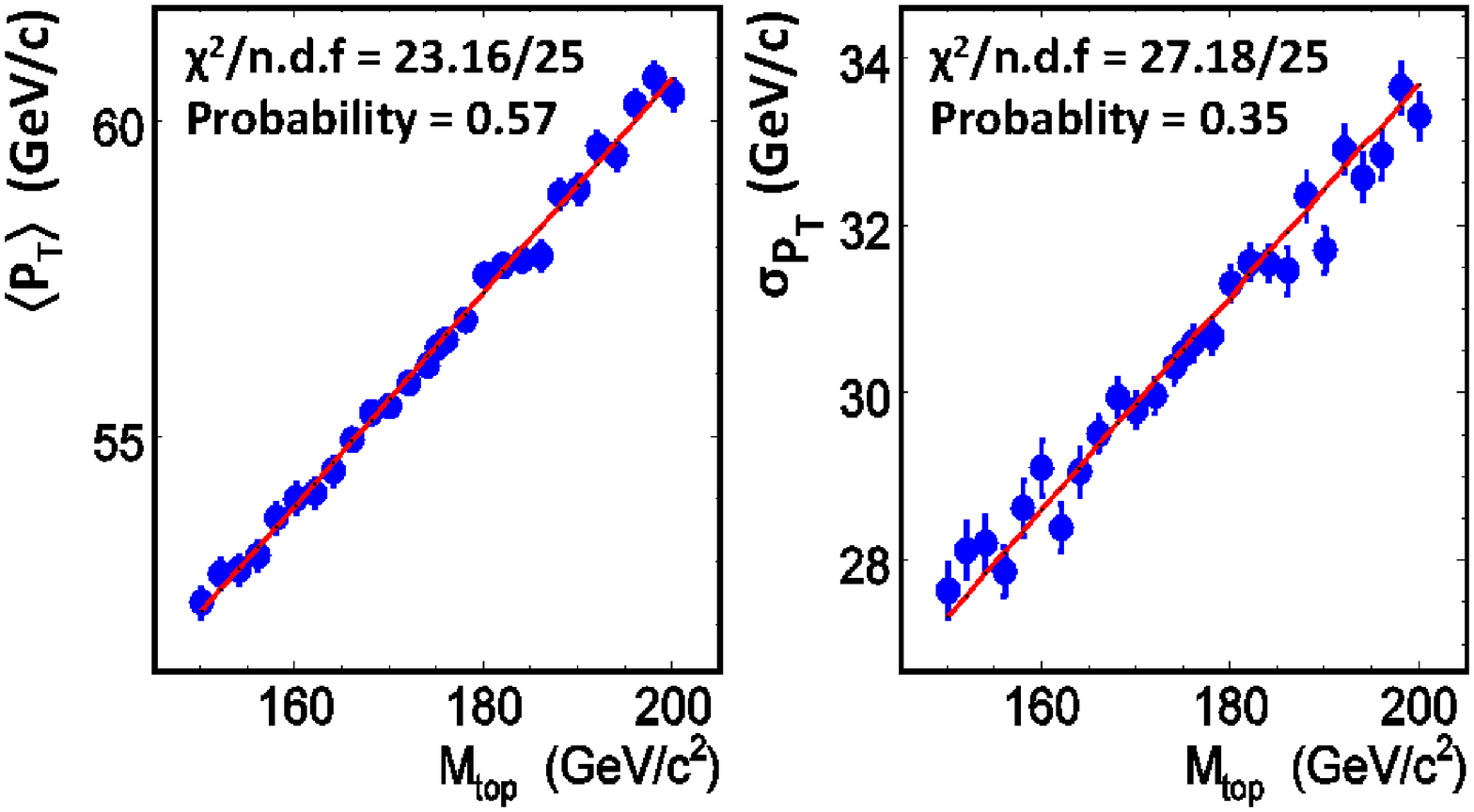}
\caption{{\bf Top:} The parameters $q$ (left) and $p$ (right) of the lepton
$P_{\rm T}$ distribution model as functions of the input $M_{top}$ of the Monte
Carlo signal templates. The plots show the linear trends of both parameters.
{\bf Bottom:} The mean (left) and standard deviation (right) of the templates
as functions of the input $M_{top}$ compared with the corresponding first two
statistical moments of the lepton $P_{\rm T}$ distribution model (straight
lines), computed using the slopes and intercepts of Table \ref{tab:coeff}.
The $\chi^{2}$ probabilities are calculated from the deviations of the points
from the lines, without fitting, and are therefore measures of validity of
the model.
\label{fig:pq}}
\end{figure}

A set of 27 PYTHIA $t\bar{t}$ samples with different central input $M_{top}$
values between 150 and 200 GeV/$c^{2}$ was analyzed. The lepton $P_{\rm T}$
histograms of the selected events from each sample were constructed for
each central $M_{top}$ input value. Rather than using directly these histograms
as binned templates to fit $M_{top}$, an analytical parametrization of the
lepton $P_{\rm T}$ distribution is chosen for this measurement. All the lepton
$P_{\rm T}$ histograms can be accurately modeled by the number of leptons $N$
in the histogram times an incomplete $\Gamma$ probability density function:

\begin{eqnarray}
\nonumber
F\left(P_{\rm T};~\! p,q\right)&=&
\frac{1/q}{\Gamma(1+p,c/q)}\left(\frac{P_{\rm T}}{q}\right)^{p}
\frac{e^{-P_{\rm T}/q}}{1+e^{(c-P_{\rm T})/a}},\\
a,~\! c,~\! p,~\! q>0,&&
\label{eq:Gamma}
\end{eqnarray}

\noindent
with two free parameters $p$ and $q$. The Fermi-Dirac factor
$1/[1+e^{(c-P_{\rm T})/a}]$ gives a finite width $a$ to the event selection
threshold at $c=20$ GeV/$c$ and tends to a unit step function
$\vartheta(c-P_{\rm T})$ in the limit of $a\to 0$ of the true, infinitely
sharp lepton $P_{\rm T}$ cut. The fit of this function to any of the lepton
$P_{\rm T}$ templates was insensitive to any choice of $a\le 0.1$ GeV/$c$,
while the $\chi^{2}$ was progressively increasing for $a>0.1$ GeV/$c$.
This parameter was thus fixed at $a=0.1$ GeV/$c$ in the incomplete $\Gamma$
function. An example of the fit of this function to the lepton $P_{\rm T}$
template with input $M_{top}=175$ GeV/$c^{2}$ is shown in the top plot of
Figure \ref{fig:fits}. The dependence of the lepton $P_{\rm T}$ distribution
on the input $M_{top}$ of the templates was studied by fitting $p$ and $q$
to each $t\bar{t}$ signal template. The fits are shown in the top plots of
Figure \ref{fig:pq}. The parameter $p$ shows significant local fluctuations
because it is mostly sensitive to the location of the distribution's maximum
which lies very close to the cut at 20 GeV/$c$. Therefore, the individual fit
to each template does not constrain this parameter strongly enough. Apart
from this the fits show an approximately linear dependence of both parameters
on $M_{top}$. Based on this observation, the dependence was modeled by leading
order Taylor expansions in terms of $M_{top}$:

\begin{equation}
p=\alpha_{1}+\alpha_{2}M_{top}~~~~~q=\alpha_{3}+\alpha_{4}M_{top},
\label{eq:shape}
\end{equation}

\noindent
where terms of $O(M_{top}^{2})$ were dropped. The zeroth and first order
coefficients $\alpha_{1,2,3,4}$ were determined from a fit of Equations
\ref{eq:Gamma} and \ref{eq:shape} to all $t\bar{t}$ signal templates
simultaneously using the program MINUIT \cite{minuit}. The simultaneous fit
smooths out local fluctuations of the parameters, giving an improved
$\chi^{2}$ probability. The results are shown in Table \ref{tab:coeff}.
Coefficient $\alpha_{1}$ is anti-correlated with $\alpha_{2}$, and $\alpha_{3}$
with $\alpha_{4}$, at the level of 60\% in either case, whereas other
correlations are much smaller. This parameterization encapsulates all of the
$M_{top}$ information that the MC signal templates provide. The incomplete
$\Gamma$ function was also found to model accurately the total background
template which is constructed by adding the lepton $P_{\rm T}$ histograms
of all background contributions, according to the expected rates of Table
\ref{tab:expSM}, and is independent of $M_{top}$. The fit of this function
to the total background template is shown in the bottom plot of Figure
\ref{fig:fits}. The background lepton $P_{\rm T}$ distribution was, therefore,
modeled using the same constants $c=20$ GeV/$c$ and $a=0.1$ GeV/$c$ and fitting
the free parameters $p$ and $q$. The background fit values $\beta_{1}$ of $p$
and $\beta_{2}$ of $q$ are also shown in Table \ref{tab:coeff}. The model was
validated by verifying that the first two statistical moments of both signal
and background template histograms were reproduced, within statistical
uncertainties, by the incomplete $\Gamma$ function using the fit parameters
of Table \ref{tab:coeff}. This is shown for the signal templates in the
bottom plots of Figure \ref{fig:pq}, where the two moments computed directly
from the templates are compared with the values obtained by integrating the
incomplete $\Gamma$ function. It is worth noting that the two moments depend
linearly on $M_{top}$ as well.

An unbinned likelihood, $L$, was constructed based on the modeling of the
lepton $P_{\rm T}$ distributions for signal and background events:

\begin{eqnarray}
\nonumber
L&=&\frac{1}{\sqrt{2\pi}~\!\delta b}
\exp\left[-~\!\frac{1}{2}\left(\frac{n_{b}-b}{\delta b}\right)^{2}\right]\\
&\times&\frac{(n_{s}+n_{b})^{N}e^{-(n_{s}+n_{b})}}{N!}\nonumber\\
&\times&\prod_{i=1}^{N}~\!
\frac{n_{s}~\!F\left(P_{\rm T}^{(i)};~\!\vec{\alpha};~\! M_{top}\right)
+n_{b}~\!F\left(P_{\rm T}^{(i)};~\!\vec{\beta}\right)}{n_{s}+n_{b}}
\label{eq:like}
\end{eqnarray}

\noindent
The likelihood contains the product of the normalized probabilities of
$n_{s}$ leptons to come from the $t\bar{t}$ signal and $n_{b}$ leptons to
come from the background. By fixing the shape parameters
$\vec{\alpha}=(\alpha_{1},\alpha_{2},\alpha_{3},\alpha_{4})$ of the signal
and $\vec{\beta}=(\beta_{1},\beta_{2})$ of the background to the values of
Table \ref{tab:coeff}, the likelihood becomes a function of $M_{top}$ and
of the numbers $n_{s}$ and $n_{b}$. It contains a Gaussian constraint
relating $n_{b}$ to the expected total number $b$ of background leptons,
with $\delta b$ being the uncertainty of this number derived by adding
quadratically the uncertainties of Table \ref{tab:expSM}, and a Poisson
constraint relating the sum $n_{s}+n_{b}$ to the number $N$ of observed
leptons of Table \ref{tab:expSM}.

\section{Corrections and tests}
\label{sec:corrections}

\begin{figure}
\includegraphics[width=8.66cm]{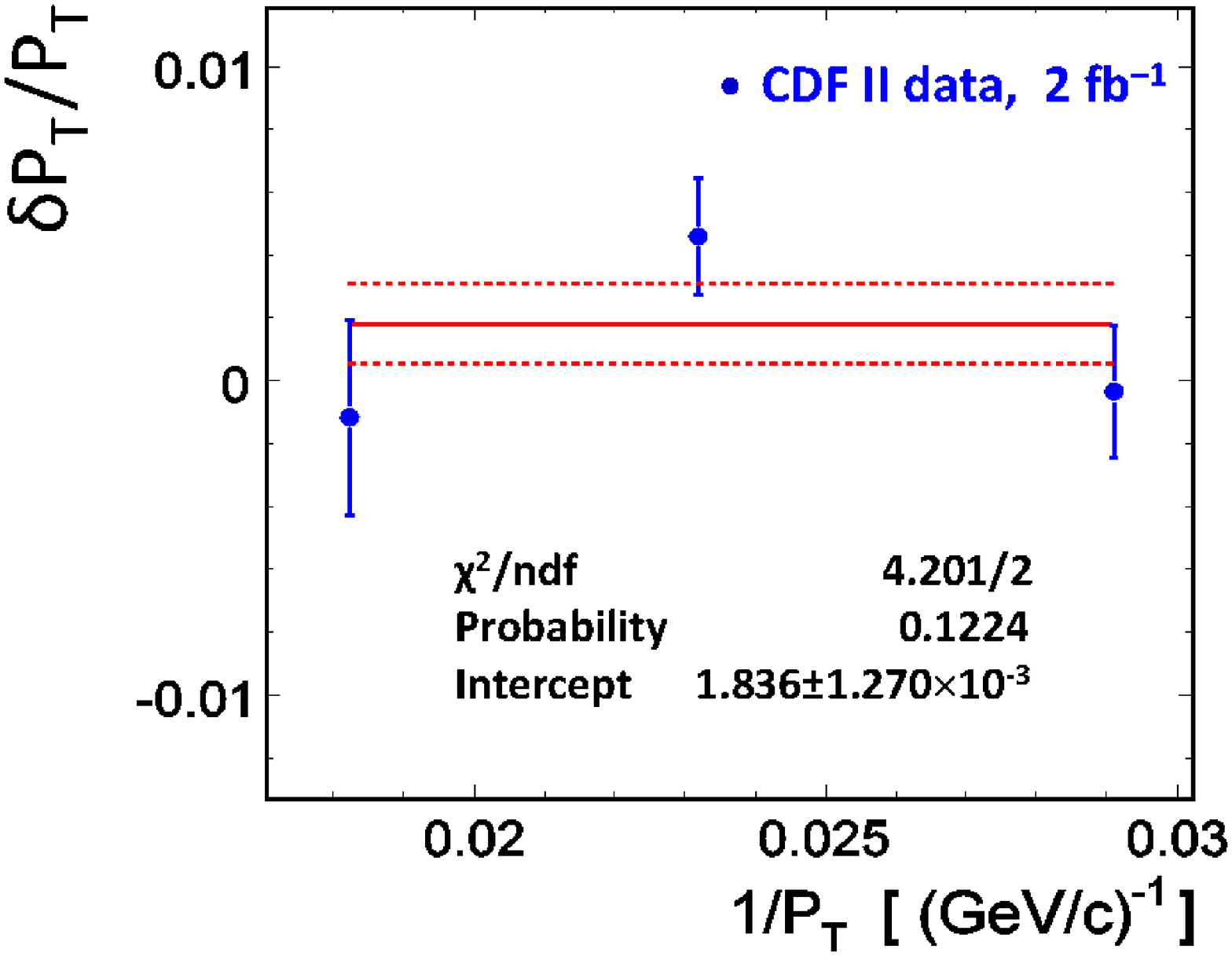}
\vspace{-0.2cm}
\includegraphics[width=8.66cm]{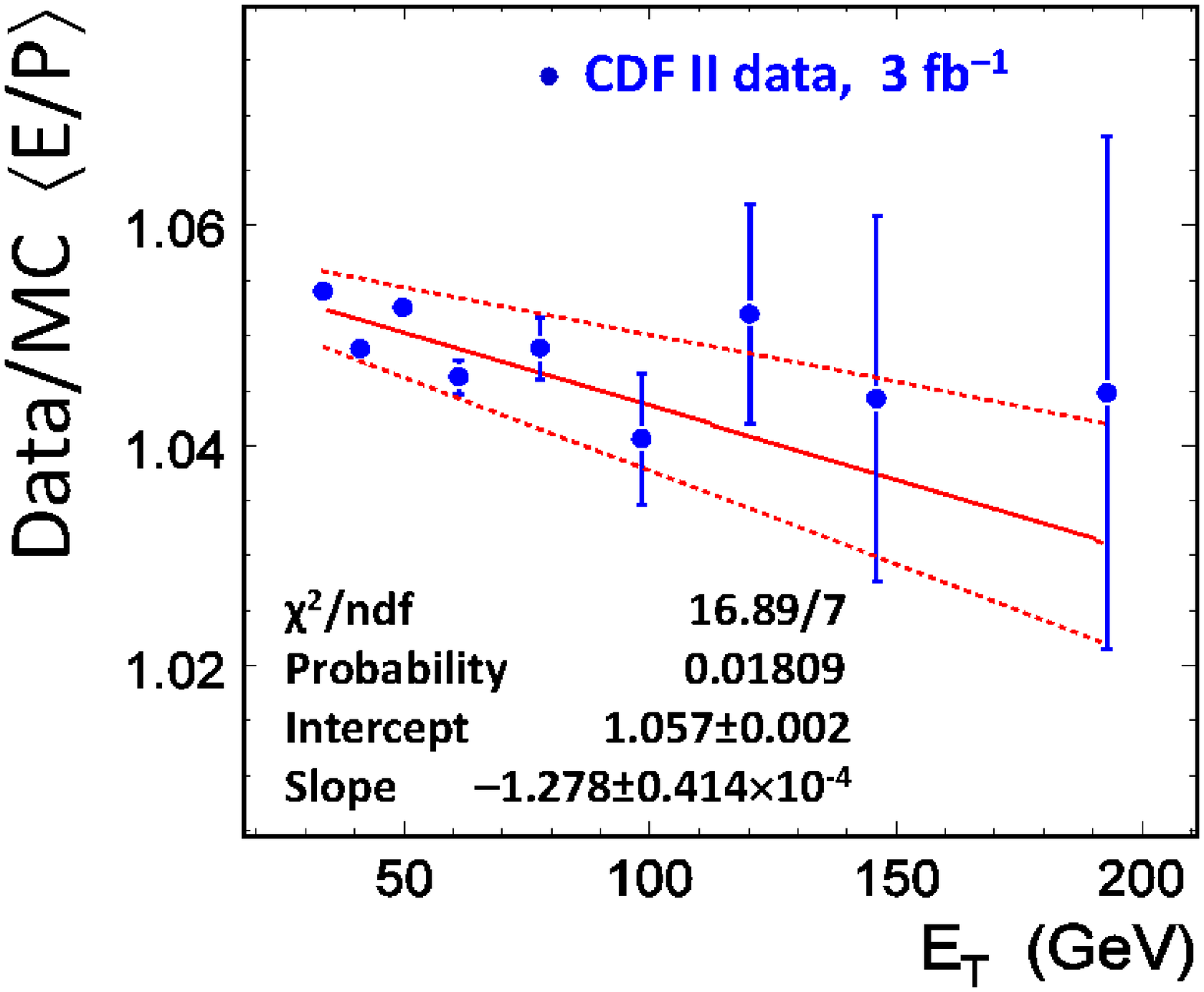}
\caption{{\bf Top:} The constant term fit of the muon relative change of
scale $\delta P_{\rm T}/P_{\rm T}$ as a function of $1/P_{\rm T}$ for
$Z\to\mu^{+}\mu^{-}$ data. {\bf Bottom:} The linear fit of the data to
MC ratio of the electron average $E/P$ over each $E_{\rm T}$ bin for $e+1$
jet data and MC. In each plot, the dashed lines represent $\pm 1\sigma$
variations of the respective fit.
\label{fig:pt}}
\end{figure}

A detailed calibration of the lepton $P_{\rm T}$ scale was performed. The
overall scale was calibrated by tuning the reconstructed $Z\to e^{+}e^{-}$
and $Z\to\mu^{+}\mu^{-}$ mass peaks of the data and MC samples to the $Z$
mass world average \cite{pdg}. The correction applied to the electron
$E_{\rm T}$ scale is $+0.4\%$ for the data and $-0.4\%$ for the MC. The
correction to the muon $P_{\rm T}$ scale is $+0.4\%$ for the data, while
no significant shift was found in the MC. The local muon $P_{\rm T}$ scale
was calibrated by binning the data and MC samples in $1/P_{\rm T}$ and
reconstructing the $Z\to\mu^{+}\mu^{-}$ mass peak in each bin. Local
changes of the $P_{\rm T}$ scale were examined by tuning to the $Z$ mass
world average. The top plot of Figure \ref{fig:pt} shows the relative change
in the scale as a function of $1/P_{\rm T}$, which is proportional to the
muon track curvature in the magnetic field of the detector. The fit of a
constant term describes the points reasonably well, showing no significant
local change in the $P_{\rm T}$ scale. The local electron $E_{\rm T}$ scale
was calibrated by correcting the slope of the energy to momentum $E/P$ ratio
as a function of $E_{\rm T}$ of electron + 1 jet data and MC samples, from
which an $E_{\rm T}$-dependent correction was derived. The $E/P$ ratio is
assumed to be insensitive to the global electron momentum scale. The $e+1$
jet sample contains $W\to e+\nu_{e}$ events associated with exactly one jet
at the level of $\sim$80\%. It was chosen for best balance between high
statistics, moderate background from jets misidentified as electrons
(``fake''electrons) and wide $E_{\rm T}$ range. The bottom plot of Figure
\ref{fig:pt} shows a linear fit of the data to MC $E/P$ ratio as a function
of $E_{\rm T}$ from which the electron $E_{\rm T}$ calibration parameters are
derived. Although moderately good, the fit suggests a decrease of the ratio
with increasing $E_{\rm T}$, which can be attributed to energy leakage in the
calorimeter. A possible effect of fake electrons on the fit was examined by
cutting on $0.8<E/P<1.2$ where the $E/P$ ratio of true electrons peaks. The
$E/P$ ratio of fake electrons is random, thus adding a flat background to
the $E/P$ spectrum. The cut, therefore, eliminates most fake electrons. No
significant difference between the results of the fit with and without the
cut was observed. In both cases of muon and electron local scale, polynomial
fits of higher degree yield coefficients of higher order terms of sizes well
within the errors, having negligible effect on the calibration relative to
the uncertainties of the fits shown in Figure \ref{fig:pt}.

Generator level comparisons of lepton $P_{\rm T}$ spectra from PYTHIA and
MC$@$NLO \cite{mcatnlo} showed that the lepton $P_{\rm T}$ distribution is
most sensitive to next-to-leading order (NLO) effects in the initial state
of $t\bar{t}$ events. The signal MC, generated with the leading order (LO)
PYTHIA generator, was thus corrected for NLO effects involving the initial
state. The $t\bar{t}$ signal events were reweighted from the LO in
$\alpha_{s}$ CTEQ5L \cite{cteq5l} set of parton distribution functions (PDF)
of the proton, which is the default in PYTHIA 6.2, to the NLO CTEQ6M set
\cite{cteq6m}. In addition, the LO 6\% fraction of $gg\to t\bar{t}$ events
of PYTHIA was scaled up to the NLO fraction of 15\% \cite{nlo}.

The robustness of the method over the full range of $M_{top}$ values covered
by the MC signal templates was tested with simulated experiments, using in
each experiment the number of events observed in the data and the expected
sample composition of Table \ref{tab:expSM}. The signal and background events
were randomly sampled for each experiment from the respective templates and
a new fit was performed each time using the parameters of Table
\ref{tab:coeff} and maximizing the likelihood defined by Equation
\ref{eq:like}. It was found in all cases that the method is unbiased and
the statistical uncertainty of the measured $M_{top}$ is correctly estimated.
The expected relative statistical uncertainty is 4.5\% after the lepton
$P_{\rm T}$ scale corrections and the reweighting from LO to NLO PDF are
applied.

\section{Result}
\label{sec:result}

\begin{figure}
\includegraphics[width=8.66cm]{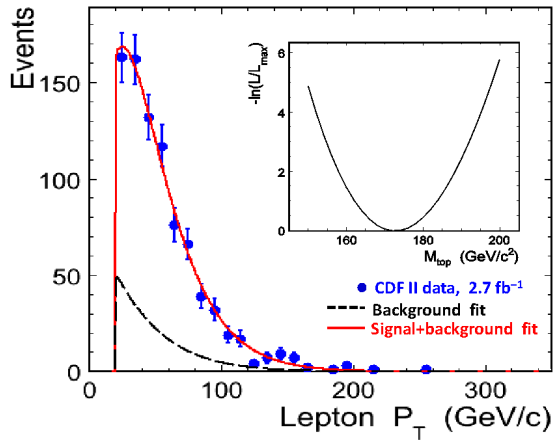}
\caption{The unbinned maximum likelihood fit of $M_{top}$ to the data. The
inset plot shows the log-likelihood curve of the fit. The result of the fit
before any correction is $M_{top}=171.9\pm 7.9_{\rm stat}$ GeV/$c^{2}$.
\label{fig:fit}}
\end{figure}

A maximum likelihood fit was performed to the 2.7 fb$^{-1}$ data sample
consisting of 858 lepton+jets events, 472 of which are electron+jets events
and 386 are muon+jets. The fit is shown in Figure \ref{fig:fit} and
the result is $M_{top}=171.9\pm 7.9_{\rm stat}$ GeV/$c^{2}$ before any
corrections. The $\chi^{2}$/n.d.f. of the fit is 21.4/27=0.79, corresponding
to a $\chi^{2}$ probability of 0.77. The total $P_{\rm T}$ scale correction
shifts the result of the fit by $+$2.6 GeV/$c^{2}$. The total correction of
the result for the NLO reweighting is $+$2.4 GeV/$c^{2}$. The two
corrections add to an overall correction of $+$5.0 GeV/$c^{2}$ of the fit
result to give a final result of $M_{top}=176.9\pm 8.0_{\rm stat}$
GeV/$c^{2}$, where the increase of 0.1 GeV/$c^{2}$ in the statistical
uncertainty follows from the increase in the central value.

All systematic uncertainties are determined by performing simulated
experiments in which the systematic parameter in question is varied, the
default method and corrections are applied, and the shift in the average
measured top quark mass with respect to the value measured from the nominal
sample is used to quantify the uncertainty. The systematic uncertainties are
summarized in Table \ref{tab:sys}. The uncertainty from the finite MC
statistics was estimated by varying the shape parameters of Table
\ref{tab:coeff} by $\pm 1\sigma$. The uncertainty from the lepton $P_{\rm T}$
scale was estimated by varying the lepton $P_{\rm T}$ correction parameters
by $\pm 1\sigma$ of the respective fit from which each parameter was derived.
This uncertainty is sizeable and almost entirely originating from the local
scale calibration where the information provided by the data is poor, as seen
in Figure \ref{fig:pt}. An estimate of the uncertainty from the choice of the
MC event generator was obtained by comparing the fit to the default PYTHIA
$t\bar{t}$ sample with the fit to a HERWIG \cite{herwig} $t\bar{t}$ sample,
including the total background in both cases. The uncertainty from the proton
PDF set was estimated by varying the CTEQ6M eigenvectors and the $\alpha_{s}$
value within their 90\% confidence level intervals. For the gluon initial
and final state radiation, an estimate of the uncertainty was obtained by
comparing the fits to two signal+background MC samples with higher and lower
radiation with the fit to the default sample. For the multiple hadron
interactions, the uncertainty was estimated by reweighting the default MC
sample to the average number of vertices in the high instantaneous luminosity
part of the data. For the background shape uncertainties, the uncertainty of
the $W$+jets component due to the choice of the $Q^{2}$ scale was estimated
by varying the $Q^{2}$ scale by a factor of 2 up and a factor of 2 down
relative to the default, and the uncertainty due to the amount of fakes
by varying the expected amount of fakes of Table \ref{tab:expSM} by
$\pm 1\sigma$, while keeping the normalization of the total background fixed.
The variation of the fakes fraction in the total background is $\sim$65\% and
affects the shape of the total background distribution to a degree consistent
with the shape discrepancy between data and expectations in the lepton + 2
jets control sample for $P_{\rm T}<50$ GeV/$c$, seen in Figure \ref{fig:bkg}.
This is the largest source of systematic uncertainty in this measurement.
Finally, an estimate of the uncertainty from the jet energy scale was
obtained by varying the combined jet energy corrections by $\pm 1\sigma$
\cite{jets} and it was found to be negligible. The total systematic
uncertainty was estimated by adding all individual uncertainties in
quadrature and was found equal to 2.7 GeV/$c^{2}$.

\begin{table}
\caption{Systematic uncertainties of the measurement. All uncertainties
are estimated at $M_{top}=175$ GeV/$c^{2}$.
\label{tab:sys}}
\begin{center}
\begin{tabular}{lc}
\hline\hline
Source & Uncertainty (GeV/$c^{2}$) \\ \hline
MC statistics & 0.4 \\
Lepton $P_{\rm T}$ scale & 1.1 \\
MC generator & 1.2 \\
Proton PDF & 0.6 \\
Gluon radiation & 0.8 \\
Multiple interactions & 0.1 \\
$Q^{2}$ scale & 0.5 \\
Fakes & 1.8 \\
JES & negligible \\ \hline
Total & 2.7 \\ \hline\hline
\end{tabular}
\end{center}
\end{table}

\section{Summary and conclusions}
\label{sec:summary}

In summary, the top quark mass has been measured using a shape analysis of
the lepton $P_{\rm T}$ distribution from a sample of 2.7 fb$^{-1}$ of CDF II
data. Events were selected in the lepton + $\ge$4 jets topology with at
least one jet tagged as coming from a $b$ quark. A MC derived model of the
dependence of the lepton $P_{\rm T}$ distribution on $M_{top}$ was used in
an unbinned maximum likelihood fit to the data. Corrections for a detailed
lepton $P_{\rm T}$ scale calibration and for NLO effects in the MC model
of $t\bar{t}$ production were applied to the fit result in order to reduce
systematic uncertainties from these two sources. Uncertainties from the jet
energy scale are negligible. The dominant uncertainty was found to come from
the shape model of the background, due to the large uncertainty in the
expected fraction of fake electrons in the selected events. The final result
is

\begin{eqnarray}
\nonumber
M_{top}=176.9\pm 8.0_{\rm stat}\pm 2.7_{\rm syst}{\rm~GeV/}c^{2}
\nonumber
\end{eqnarray}

\noindent
in good agreement, within errors, with the current world average
\cite{world}.

Compared with the previous measurement of exploiting the mean value of the
lepton $P_{\rm T}$ with data corresponding to 1.9 fb$^{-1}$ integrated
luminosity in the lepton+jets channel \cite{Lxy}, the new result shows an
appreciable reduction, from 3.8 GeV/$c^{2}$ to 2.7 GeV/$c^{2}$, in the total
systematic uncertainty. This is achieved by the use of the shape information
of the lepton $P_{\rm T}$ distribution which is less sensitive to acceptance
related effects that can change the mean $P_{\rm T}$ without altering
significantly the shape of the $P_{\rm T}$ spectrum, such as the JES and
multiple interactions, and by the new lepton $P_{\rm T}$ calibration, which
reduced the $P_{\rm T}$ scale uncertainties.

\begin{acknowledgments}
We thank the Fermilab staff and the technical staffs of the participating
institutions for their vital contributions. This work was supported by the
U.S. Department of Energy and National Science Foundation; the Italian
Istituto Nazionale di Fisica Nucleare; the Ministry of Education, Culture,
Sports, Science and Technology of Japan; the Natural Sciences and Engineering
Research Council of Canada; the National Science Council of the Republic of
China; the Swiss National Science Foundation; the A.P. Sloan Foundation;
the Bundesministerium f\"ur Bildung und Forschung, Germany; the Korean World
Class University Program, the National Research Foundation of Korea; the
Science and Technology Facilities Council and the Royal Society, UK; the
Institut National de Physique Nucleaire et Physique des Particules/CNRS;
the Russian Foundation for Basic Research; the Ministerio de Ciencia e
Innovaci\'{o}n, and Programa Consolider-Ingenio 2010, Spain; the Slovak
R\&D Agency; and the Academy of Finland. 
\end{acknowledgments}


\begin{thebibliography}{6}
\bibitem{frame}
The transverse momentum $P_{\rm T}=P\sin\theta$ and transverse energy
$E_{\rm T}=E\sin\theta$ of a particle are defined from its momentum $P$
or energy $E$, respectively, and the polar angle $\theta$ of its momentum
vector in a system of spherical coordinates with the polar axis along
the proton beam. The pseudorapidity of the particle is given by
$\eta=-\ln[\tan(\theta/2)]$.
\bibitem{met}
The missing transverse energy, measuring the transverse energy inbalance
of the event, is defined by ${\not\!\! E}_{\rm T}=\vert\sum_{towers}
{E}_{\rm T}\hat{n}_{\rm T}\vert$,
where $\hat{n}_{\rm T}$ is the unit vector normal to the beam and pointing
to a given calorimeter tower and ${E}_{\rm T}$ is the transverse energy
measured in that tower.
\bibitem{iso}
The calorimeter isolation is defined as the difference of the total
$E_{\rm T}$ in a cone of radius
$R=\sqrt{(\Delta\eta)^{2}+(\Delta\phi)^{2}}=0.4$ around the axis of a tower
cluster minus the total $E_{\rm T}$ in the cluster, where $\phi$ is the
azimuth in the spherical coordinate system.
\end{thebibliography}
\end{document}